\documentclass{article}

\usepackage[backend=biber, style = nature, citestyle = nature, sorting=none]{biblatex}
\usepackage{amssymb}
\usepackage{csquotes}

\usepackage[letterpaper,top=2cm,bottom=2cm,left=3cm,right=3cm,marginparwidth=1.75cm]{geometry}

\usepackage{amsmath}
\usepackage{graphicx}

\usepackage{hyperref}
\hypersetup{colorlinks,allcolors=blue}
\usepackage{subcaption,booktabs}
\usepackage{xcolor,colortbl}
\definecolor{lavender}{rgb}{0.9, 0.9, 0.98}

\usepackage{siunitx}
\bibliography{main.bib}

\title{Spin-Orbit Coupling for Optical Vortex Generation in van der Waals Materials}

\usepackage{authblk}

\author[1,$\dagger$]{Jaegang Jo}
\author[2,$\dagger$]{Sujeong Byun}
\author[1]{Munseong Bae}
\author[3]{Jianwei Wang}
\author[1,a,*]{Haejun Chung}
\author[2,b,*]{Sejeong Kim}

\affil[1]{Department of Electronic Engineering, Hanyang University, Seoul, 04763, Republic of Korea}
\affil[2]{Department of Electrical Engineering, Faculty of Engineering and Information Technology, University of Melbourne, Melbourne 3000, Australia}
\affil[3]{State Key Laboratory for Mesoscopic Physics and Collaborative Innovation Center of Quantum Matter, School of Physics, Peking University, Beijing, 100871, China
}
\affil[$\dagger$]{These authors contributed equally to this work.}
\affil[$*$]{These authors are corresponding authors.}

\usepackage{setspace}
\begin{document}
\maketitle

\begin{center}
    $^\text{a}$ haejun@hanyang.ac.kr 
    $^\text{b}$ sejeong.kim@unimelb.edu.au
\end{center}

\vspace{1pc}

\begin{abstract}
An optical vortex beam has attracted significant attention across diverse applications, including optical manipulation, phase-contrast microscopy, optical communication, and quantum photonics. To utilize vortex generators for integrated photonics, researchers have developed ultra-compact vortex generators using fork gratings, metasurfaces, and integrated microcombs. However, those devices depend on costly, time-consuming nanofabrication and are constrained by the low signal-to-noise ratio due to the fabrication error. As an alternative maneuver, spin-orbit coupling has emerged as a method to obtain the vortex beam by converting spin angular momentum (SAM) without nanostructures. Here, we demonstrate the creation of an optical vortex beam using van der Waals (vdW) materials. The significantly high birefringence of vdW materials allows generations of optical vortex beams with high efficiency in a sub-wavelength thickness. In this work, we utilize an 8-\unit{\micro\metre}-thick hexagonal boron nitride (hBN) crystal for the creation of optical vortices carrying topological charges of $\pm$2. We also present the generation of an optical vortex beam in a 320-nm-thick MoS$_2$ crystal with a conversion efficiency of 0.09. This study paves the way for fabrication-less and ultra-compact optical vortex generators, which can be applied for integrated photonics and large-scale vortex generator arrays.
\end{abstract}
\vspace{2pc}

%
%
%

\section{Introduction}
Over the past decades, optical vortex beams, characterized by their orbital angular momentum (OAM), have been extensively investigated for various applications, including optical micro-manipulation~\cite{da2024tailoring, lokesh2021realization, wen2022precise, huang2023metasurface}, chirality sensing~\cite{sakamoto2021chirogenesis, brullot2016resolving, forbes2021optical}, and phase contrast microscopy~\cite{huo2020photonic, furhapter2005spiral}. Furthermore, the number of topological charges, defined by the number of 2$\pi$ phase shifts along the beam's axis, allows an enhanced degree of freedom and increased data capacity for communication~\cite{yang2020generating, white2022inverse, krenn2016twisted}. This degree of freedom can also be used to multiplex states of qubits and realize OAM qubits, opening exciting research avenues in quantum optics~\cite{cozzolino2019orbital, tang2020harmonic, wu2023optical, de2024nonlinear}.

 Since the concept of optical vortices was introduced in 1989~\cite{coullet1989optical}, extensive studies have been dedicated to developing methods for generating orbital angular momentum (OAM). Over the years, various commercially available solutions have successfully demonstrated the creation of optical vortices with various topological charges. These solutions include spiral phase plates~\cite{beijersbergen1994helical, wei20193d}, q-plates~\cite{slussarenko2011tunable}, and computer-generated holograms, also known as spatial light modulators~\cite{carpentier2008making, vijayakumar2019generation}. While these methods are reliable in producing vortex beams, they often involve bulky optical systems. Consequently, there has been significant interest in advancing more compact and integrated photonics solutions for optical vortex generation. Early approaches involved the use of liquid crystals, which offered some level of integration.~\cite{son2014optical} Recent advancements have introduced the use of fork gratings~\cite{zheng2023versatile, yang2023highly}, metasurfaces~\cite{mei2023cascaded, zhao2013metamaterials, liu2021multifunctional, yu2011light}, integrated microcombs~\cite{chen2024highly, chen2024integrated, liu2024integrated, cai2012integrated}, and inverse designed devices~\cite{white2022inverse, bae2023inverse}, which are capable of generating optical vortices on a chip. However, existing on-chip solutions involve nano-scale features that require costly and time-consuming nano-fabrications and often encounter challenges in achieving adequate signal-to-noise ratios. 

While previous on-chip vortex generation methods usually involve phase-front modulation to convert non-structured light into helical light, an alternative solution has emerged suggesting the possibility of creating optical vortices by harnessing optical spin-orbit coupling.~\cite{bliokh2015spin, guo2016merging} A theoretical proof in the previous study~\cite{ciattoni2003circularly} demonstrated that when circularly polarized light passes through an uniaxial medium, a portion of the incoming light converts its handedness, thus imparting OAM to the transmitted beam. Later, this theoretical concept has been experimentally validated using lithium niobate (LN)~\cite{song2022topological}, and beta-barium borate (BBO) crystals~\cite{tang2020harmonic, wu2023optical}. However, these materials have small birefringence, which are 0.09 and 0.12  at $\lambda = 594$ nm for LN~\cite{zelmon1997infrared}, and BBO crystals~\cite{eimerl1987optical} respectively. Thus, the vortex generators utilizing these crystals necessitate bulky crystals, often exceeding several millimeters in size, to achieve high conversion efficiencies~\cite{brasselet2009dynamics}.

Here, we propose and demonstrate a fabrication-free optical vortex generator based on van der Waals (vdW) materials. By leveraging the giant birefringence in vdW materials, we efficiently induce spin-orbit coupling within the medium to generate an optical vortex beam. For this demonstration, we selected hexagonal boron nitride (hBN) and molybdenum disulfide (MoS$_2$) among various vdW materials due to their high birefringence. hBN is transparent across a wide range, from visible to infrared, while exhibiting high birefringence ($n_o \sim 2.15$ and $n_e \sim 1.86$ at $\lambda = 594$ nm)~\cite{rah2019optical}. Additionally, MoS$_2$ is known to possess extremely high optical anisotropy ($n_o \sim 4.7$ and $n_e \sim 2.7$ at $\lambda = 750$ nm) while being transparent from far-red to infrared~\cite{ermolaev2021giant}. Such large birefringence reduces the propagation length required to observe optical vortex generation through efficient spin-orbit conversion. 

We experimentally demonstrate the generation of the optical vortex beam using an 8-\unit{\micro\metre}-thick hBN crystal slab. The topological charge of the vortex is verified through an interferometer setup and also by numerical simulations. In addition, we experimentally obtain spin-orbit conversion efficiencies of vortex generators and compare them with analytical predictions. The 8-\unit{\micro\metre}-thick hBN crystal exhibits a conversion efficiency of 0.30 when a 594 nm laser is focused by an objective lens with 0.9 numerical aperture (NA). In addition, with the 26-\unit{\micro\metre}-thick MoS$_2$ crystal, we demonstrate a conversion efficiency of 0.46, which is close to the theoretical maximum value of 0.5.  

The thickness of the vortex generator can be further reduced to the sub-wavelength range. We observed the spin-orbit coupling in the 320-nm-thick MoS$_2$ crystal flake showing the conversion efficiency of 0.09. Finally, our simulations indicate that using a Bessel beam can further reduce the generator's thickness and realize near-unity conversion efficiency. This work realizes the first application of spin-orbit coupling in vdW materials for generating optical vortex beams, opening an exciting avenue to create fabrication-free and versatile optical vortex generators feasible to both small- and large-scale experiments.

\section{Results and Discussion}

\begin{figure}
    \centering
    \includegraphics[width=1.0\linewidth]{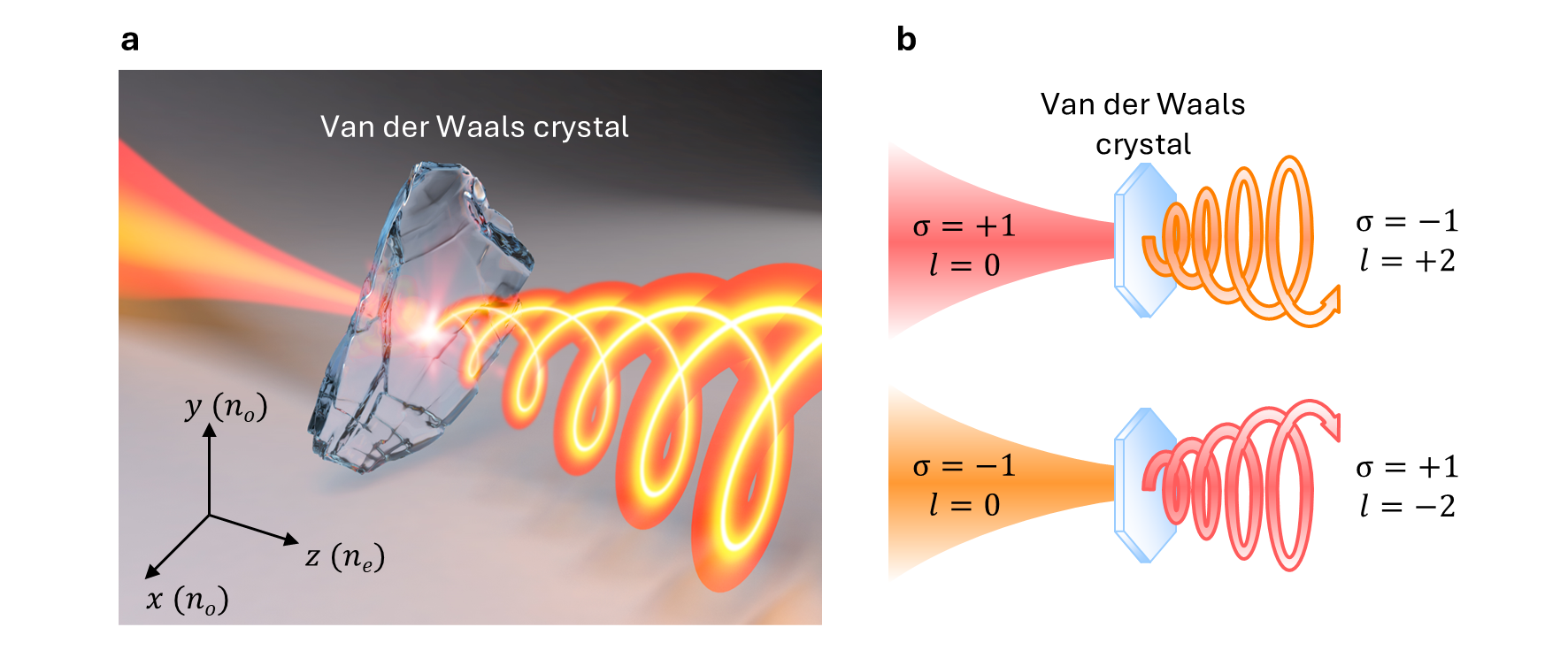}
    \caption{(a) A schematic illustration of the optical vortex generation due to spin-orbit coupling in a vdW crystal. (b) (Top) left-handed circularly polarized (LCP) light ($\sigma=+1$), initially without orbital angular momentum (OAM) ($l=0$), is incident onto a vdW crystal along its extraordinary axis, i.e. $z$-axis. As the beam propagates through the material, it is converted into a right-handed circularly polarized (RCP) beam ($\sigma=-1$) with an OAM mode of $+2$. (Bottom) When an RCP beam is incident, vdW crystal converts it into an LCP beam with an OAM mode of $-2$ ($l=-2$). }
    \label{fig:fig1}
\end{figure}

Figure~\ref{fig:fig1}a schematically illustrates the creation of optical vortices via spin-orbit coupling in the hBN crystal. As shown in Figure~\ref{fig:fig1}b, when an LCP beam ($\sigma=+1$) with zero topological charges ($l=0$) is focused on the surface of a vdW crystal and propagates along the extraordinary axis, a portion of the incident beam is converted into an RCP vortex beam ($\sigma=-1$, $l=+2$) due to the spin-orbit coupling. Conversely, if an RCP beam is incident, it generates an LCP beam with an OAM mode of $l=-2$. This conversion is proved by rigorous full-wave analysis. The time-harmonic Maxwell's equation of the complex electric field $\textbf{E}$ is expressed as $\nabla^2\textbf{E}-\nabla(\nabla\cdot \textbf{E})+k_0^2\varepsilon \cdot \textbf{E}$, where $k_0=2\pi/\lambda$. When we set the $z$-axis as the extraordinary axis of the uniaxial medium, a permittivity tensor $\varepsilon$ can be written as follows:
\begin{equation}
\varepsilon = 
    \begin{bmatrix} 
        n_o^2 & 0 & 0 \\
        0 & n_o^2 & 0 \\
        0 & 0 & n_e^2 
    \end{bmatrix},
    \label{eqn:eq1}
\end{equation}
where $n_o$ and $n_e$ are ordinary and extraordinary refractive indices. Then, the electric field propagating in the $z$-direction can be derived by Fourier transforming the ordinary (o) and extraordinary (e) plane waves in the momentum space as
\begin{equation}
    \textbf{E}(\textbf{r}_\perp, z) = \textbf{E}_o(\textbf{r}_\perp, z) +\textbf{E}_e(\textbf{r}_\perp, z) =  \iint d^2 k_\perp e^{i\textbf{k}_\perp \cdot\textbf{r}_\perp} \left[ \tilde{u}_o(\textbf{k}_\perp)e^{ik_{oz}z} {\hat{\textbf{v}}}_o(\textbf{k}_\perp) + \tilde{u}_e(\textbf{k}_\perp)e^{ik_{ez}z} \hat{\textbf{v}}_e(\textbf{k}_\perp)  \right],
        \label{eqn:eq2}
\end{equation}
where $\textbf{k}_\perp = k_x\hat{\textbf{x}}+k_y\hat{\textbf{y}}$ and $\textbf{r}_\perp = x\hat{\textbf{x}}+y\hat{\textbf{y}}$ are the transverse wavevector and position vector, respectively; and $\tilde{u}_{o,e}$ are the amplitudes of o- and e-plane wave modes. $k_{oz}=(k_0^2 n_o^2-k_\perp^2)^{1/2}$ and $k_{ez}=(k_0^2 n_e^2-k_\perp^2)^{1/2}n_o/n_e$ are the $z$-directional wavevectors of the o- and e-waves, respectively. $\hat{\textbf{v}}_{o} = -\sin\phi\hat{\textbf{x}}+\cos\phi\hat{\textbf{y}}$ and $\hat{\textbf{v}}_{e} = \left(k_{ez}/k_e n_o^2 \right)\left(\textbf{k}_\perp/k_\perp\right) - \left(k_\perp/k_e n_e^2 \right)\hat{\textbf{z}}$ are the unit vectors in the direction of the electric field of the o- and e-waves, and $\phi$ is the azimuthal angle of the plane wave's Poynting vector. The electric field can be converted into the linear combination of LCP ($+$) and RCP ($-$) electric fields as 
\begin{equation}
    \textbf{E}(\textbf{r}_\perp, z) = \textbf{E}_{+} (\textbf{r}_\perp,z) +\textbf{E}_{-} (\textbf{r}_\perp,z) = \iint d^2k_\perp e^{i\textbf{k}_\perp\cdot \textbf{r}} \left[ \tilde{U}_{+} (\textbf{k}_\perp,z)\hat{\textbf{V}}_{+} + \tilde{U}_{-} (\textbf{k}_\perp,z)\hat{\textbf{V}}_{-} \right].
    \label{eqn:eq3}
\end{equation}
Here, $\hat{\textbf{V}}_{\pm}\equiv (\hat{\textbf{x}}\pm i\hat{\textbf{y}})/\sqrt{2}$ are the unit vectors in the directions of the electric fields of the circularly polarized waves; and $\tilde{U}_\pm(\textbf{k}_\perp, z)$ are the amplitudes of the circularly polarized waves at propagation length z. The amplitudes can be derived by $\tilde{U}_\pm(\textbf{k}_\perp, 0)$ as
\begin{equation}
    \begin{bmatrix}
        \tilde{U}_+(\textbf{k}_\perp,z) \\
        \tilde{U}_-(\textbf{k}_\perp,z)
    \end{bmatrix} = 
    \begin{bmatrix}
        t_{++}(k_\perp,z) & t_{+-}(k_\perp,z) \exp(-i2\phi) \\
        t_{-+}(k_\perp,z) \exp(i2\phi) & t_{--}(k_\perp,z)
    \end{bmatrix}
    \begin{bmatrix}
        \tilde{U}_+(\textbf{k}_\perp,0) \\
        \tilde{U}_-(\textbf{k}_\perp,0)
    \end{bmatrix}.
    \label{eqn:eq4}
\end{equation}
Here, $t_{++}=t_{--} = [\exp(ik_{ez}z)+\exp(ik_{oz}z)]/2$ and $t_{+-}=t_{-+} = [\exp(ik_{ez}z)-\exp(ik_{oz}z)]/2$. The relation between the spin and orbital angular momentum can be derived from Eq.~\ref{eqn:eq3}. If an LCP beam with a topological charge $l$ is incident, $\tilde{U}_{+}(\textbf{k}_\perp,0) =  \tilde{U}_{+}(k_\perp,0)\exp(il\phi)$ and $\tilde{U}_{-}(\textbf{k}_\perp,0) = 0$. Then, the amplitudes at propagation length $z$ are
$\tilde{U}_{+}(\textbf{k}_\perp,z) = t_{++}(k_\perp , z)\tilde{U}_{+}(k_\perp,0)\exp(il\phi)$ and 
$\tilde{U}_{-}(\textbf{k}_\perp,z) = t_{-+}(k_\perp , z) \tilde{U}_{+}(k_\perp,0) \exp(i(l+2)\phi)$.
This implies that the LCP beam with the OAM mode of $l$ transfers its power to the RCP beam with the OAM mode of $l+2$ as it propagates, following the angular momentum conservation.

\begin{figure}
    \centering
    \includegraphics[width=1.0\linewidth]{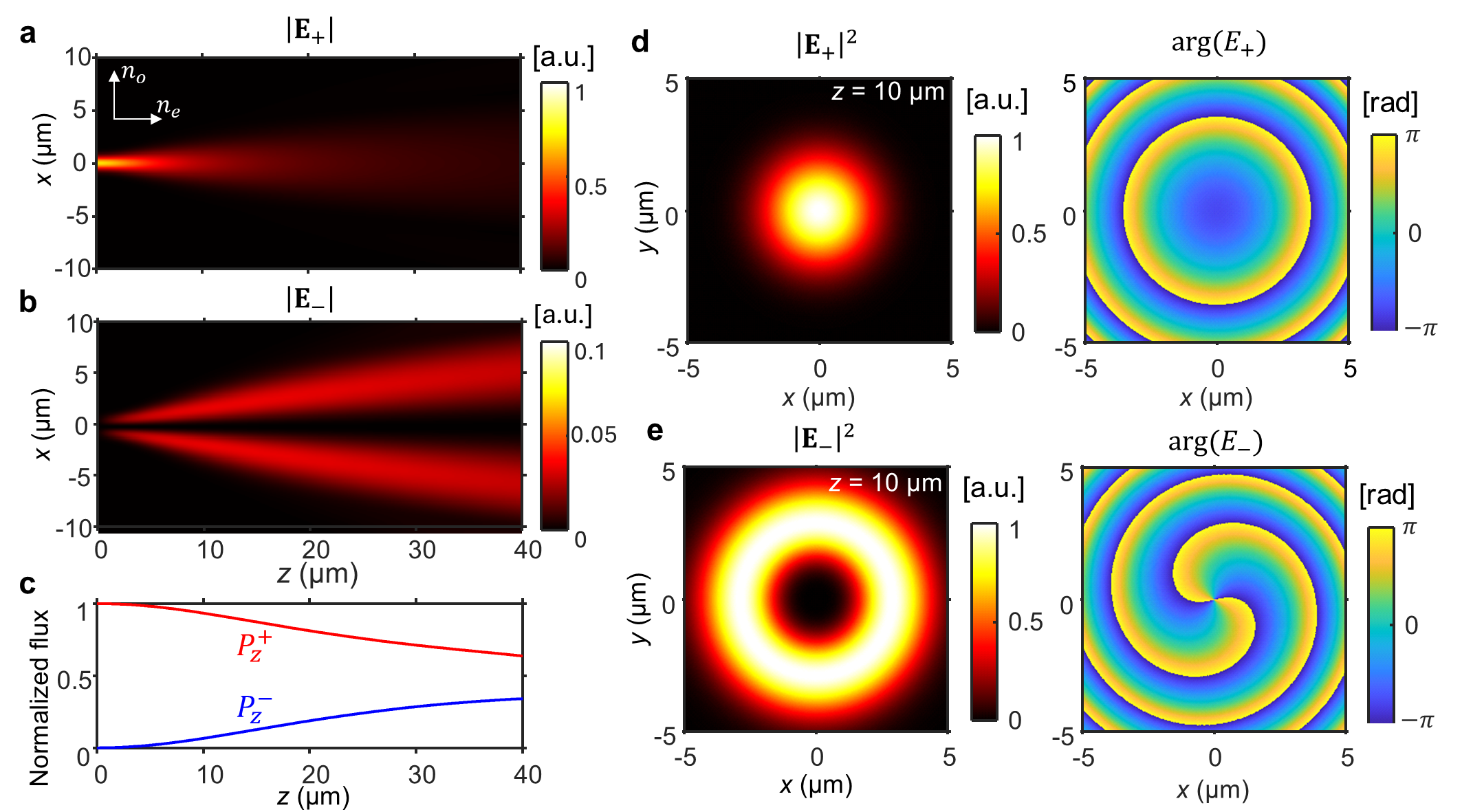}
    \caption{Numerical calculation of the spin-orbit coupling. 
    (a and b) Amplitude distributions of the LCP and RCP fields under the incidence of an LCP Gaussian beam with $\lambda = 594$ nm and a half-beam waist of 0.62 \unit{\micro\metre} propagating along the extraordinary axis of the hBN crystal ($n_o = 2.15$, $n_e = 1.86$), respectively.
    $|\textbf{E}_+|$ and $|\textbf{E}_-|$ indicate the absolute amplitudes of the LCP and RCP electric fields, respectively, normalized by the maximum amplitudes of the LCP field. 
    (c) $z$-directional powers of the LCP ($P_z^+$) and RCP ($P_z^-$) waves in the hBN crystal. (d and e) Intensity (left) and phase profiles (right) of the LCP and RCP fields at $z = 10$ \unit{\micro\metre} in the hBN crystal, respectively. The intensity profiles are normalized by their maximum intensities.
    }
    \label{fig:fig2}
\end{figure}

We conducted the cylindrical finite-difference time-domain (FDTD) simulation using open-source software MEEP~\cite{oskooi2010meep} to analyze the optical vortex generation in detail. Figure~\ref{fig:fig2}a and b show the electric field amplitude profiles of the LCP and RCP waves, respectively, illustrating the propagation of the LCP Gaussian beam in the hBN crystal ($n_o = 2.15$, $n_e = 1.86$). The light source with a wavelength of 594 nm has a half-beam waist of 0.62 \unit{\micro\metre} and propagates along the $z$-axis. Initially, the RCP intensity is zero at $z=0$ as shown in Figure~\ref{fig:fig2}b; it arises as the beam propagates along the crystal. Figure~\ref{fig:fig2}c depicts the $z$-directional power of the LCP wave $P^+_z$, being transferred to the power of the RCP wave $P^-_z$ as the propagation length $z$ increases.  Figures~\ref{fig:fig2}d and e show the intensity and phase profiles of the LCP and RCP waves, respectively. While the LCP wave maintains its Gaussian beam profile, the RCP wave displays doughnut-shaped intensity and a phase singularity at the center of the beam. In addition, the RCP wave displays a $4\pi$ phase shift in the $\phi$-direction, indicating the OAM number of $+2$.

\begin{figure}
    \centering
    \includegraphics[width=1.0\linewidth]{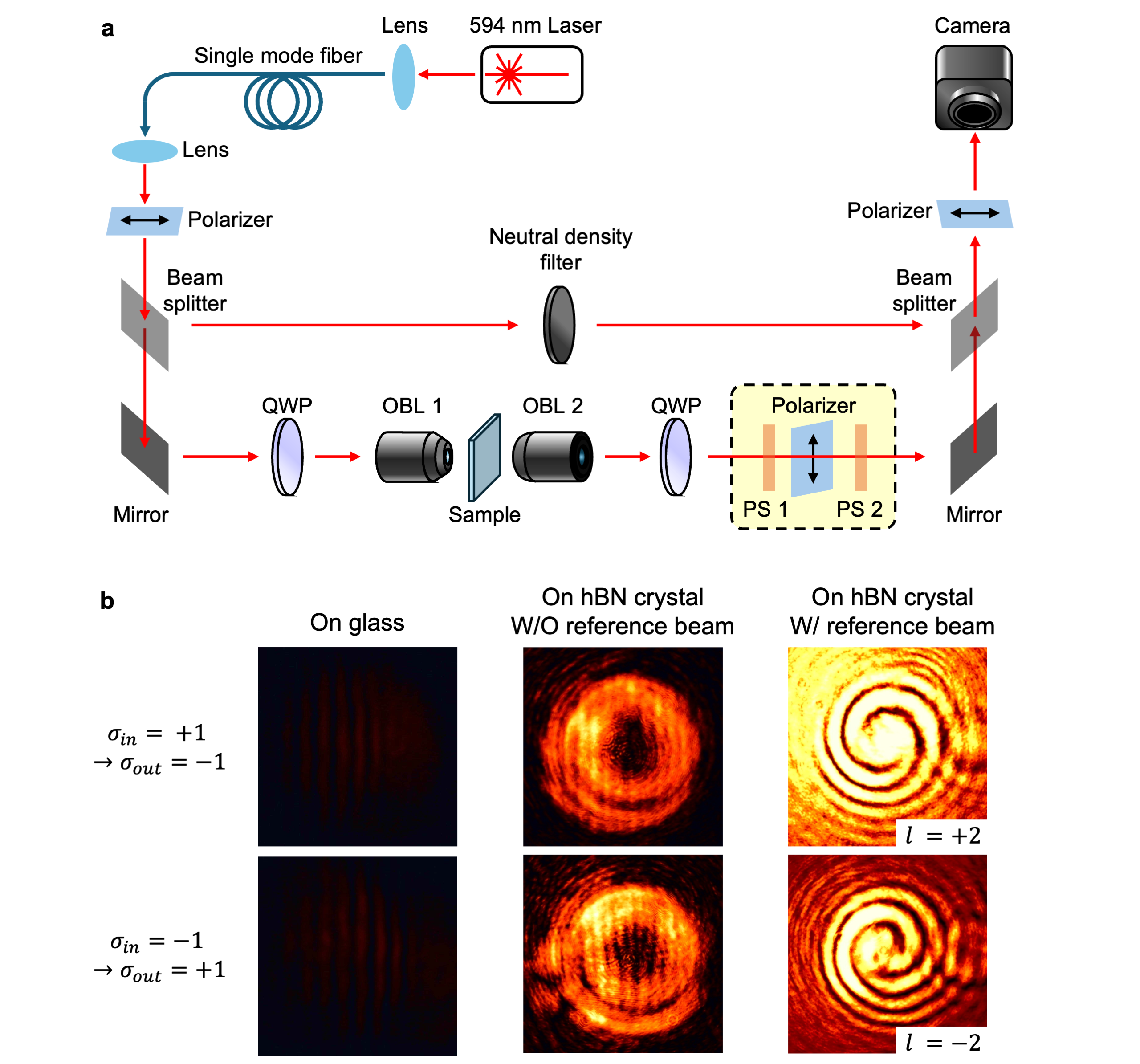}
    \caption{(a) The experimental setup. The input beam toward the sample is filtered to have left or right circular polarization. The output beams from the sample with left and right circular polarizations are converted to vertically and horizontally polarized lights by the QWP, and only horizontally polarized light passes through the linear polarizer and propagates to the camera. Two beam splitters are used to make an interference pattern of the signal and reference beams. The dashed yellow box with power sensors (PSs) indicates the optical components for measuring the conversion efficiency, which is not used for the intensity profile measurements. (b) $\sigma_{in}$ and $\sigma_{out}$ indicate the SAM number ($+1$ for LCP, $-1$ for RCP) of the input and output beams, respectively. (Left and middle columns) Intensity profiles of the output beams when the input beam focuses the bare glass substrate or the hBN crystal on the substrate, respectively. (Right column) Interference patterns of the output signal beams and the reference beam when the beam propagates both the hBN crystal on the substrate. The numerical aperture of the objective lenses (OBL) used to obtain the intensity profiles, OBL1 and OBL2, are 0.4 and 0.42, respectively.}
    \label{fig:fig3}
\end{figure}

Next, we experimentally verify the vortex generation in the vdW materials and confirm its topological charge number. Figure~\ref{fig:fig3}a shows the interferometer setup we used for the vortex beam characterization. We use a single-mode fiber and lenses to obtain a collimated Gaussian beam from a 594 nm laser. The beam passes through a linear polarizer (LP) and a quarter wave plate (QWP) to generate an LCP or RCP beam incident on the sample. The circularly polarized beam is subsequently focused by the first objective lens (OBL1), passes through the hBN crystal, and is collected by the second objective lens (OBL2). The QWP behind the OBL2 converts LCP and RCP beams into two orthogonally polarized beams, respectively. Finally, only the beam carrying the OAM is selected to pass through the LP and is observed by the camera. Additionally, two beam splitters are used to obtain the interference pattern between the OAM and the reference beam.

Figure~\ref{fig:fig3}b shows the intensity profiles when the input and output beam have opposite handedness. The images in the upper and lower rows represent the data with the input beam of LCP and RCP, respectively. When the circularly polarized light (CPL) beam passes through the hBN crystal placed on a substrate, doughnut-like intensity profiles are observed (middle column) with the opposite handedness.  In contrast, negligible intensity signals are detected (left column) when there is no hBN crystal, i.e., the incident beam is only passing through the glass substrate. This is because the polarizer before the camera selectively observes the CPL with the opposite handedness. This result shows that the hBN crystal converts the spin state and generates the annular intensity profile as the characteristic of an optical vortex beam. Next, the doughnut-shaped beams are interfered with a reference beam, resulting in two-armed spiral patterns (right column) in the interference image. These spiral arms rotate clockwise or counterclockwise depending on the incident beam, indicating the opposite sign of the topological charge of the OAM beams. These results align with the simulation results in Fig.~\ref{fig:fig2}.

In the next step, the conversion efficiency of the optical spin-orbit coupling, defined by the ratio of the output vortex beam's power to the total output power, is predicted analytically~\cite{ciattoni2003circularly, ciattoni2001vectorial, ling2020vortex}. First, the electric field of the LCP Gaussian beam is expressed as $\textbf{E}(\textbf{r}_\perp, 0)= \exp\left( -r^2_\perp / w_0^2 \right) \hat{\textbf{V}}_{+}$, where $w_0$ is half of the minimum beam waist. The amplitudes of this electric field in the Fourier space are given by $\tilde{U}_{+}(\textbf{k}_{\perp}, 0) = \left(w_0^2/2\right)\exp\left[ -(w_0 k_\perp)^2 /4 \right]$ and $\tilde{U}_{-}(\textbf{k}_{\perp}, 0) = 0$. The electric fields of the LCP and RCP waves at the propagation length $z$ are then calculated by applying Eq.~\ref{eqn:eq4}. Then, the conversion efficiency, defined as the power of the RCP wave in the $z$-direction over the total power in the $z$-direction, is approximated as follows:
\begin{equation}
    \eta = \frac{1}{2}\left[1-\frac{1}{1+(z/L)^2}\right],
    \label{eqn:eq5}
\end{equation}
where, $L=k_0n_o w_0^2/(n_o^2/n_e^2-1)$ is called the anisotropic diffraction length~\cite{ciattoni2003circularly}.
Equation~\ref{eqn:eq5} implies that the conversion efficiency depends on only $z$ and $L$, and its theoretical maximum value is 0.5.

\begin{figure}
    \centering
    \includegraphics[width=1\linewidth]{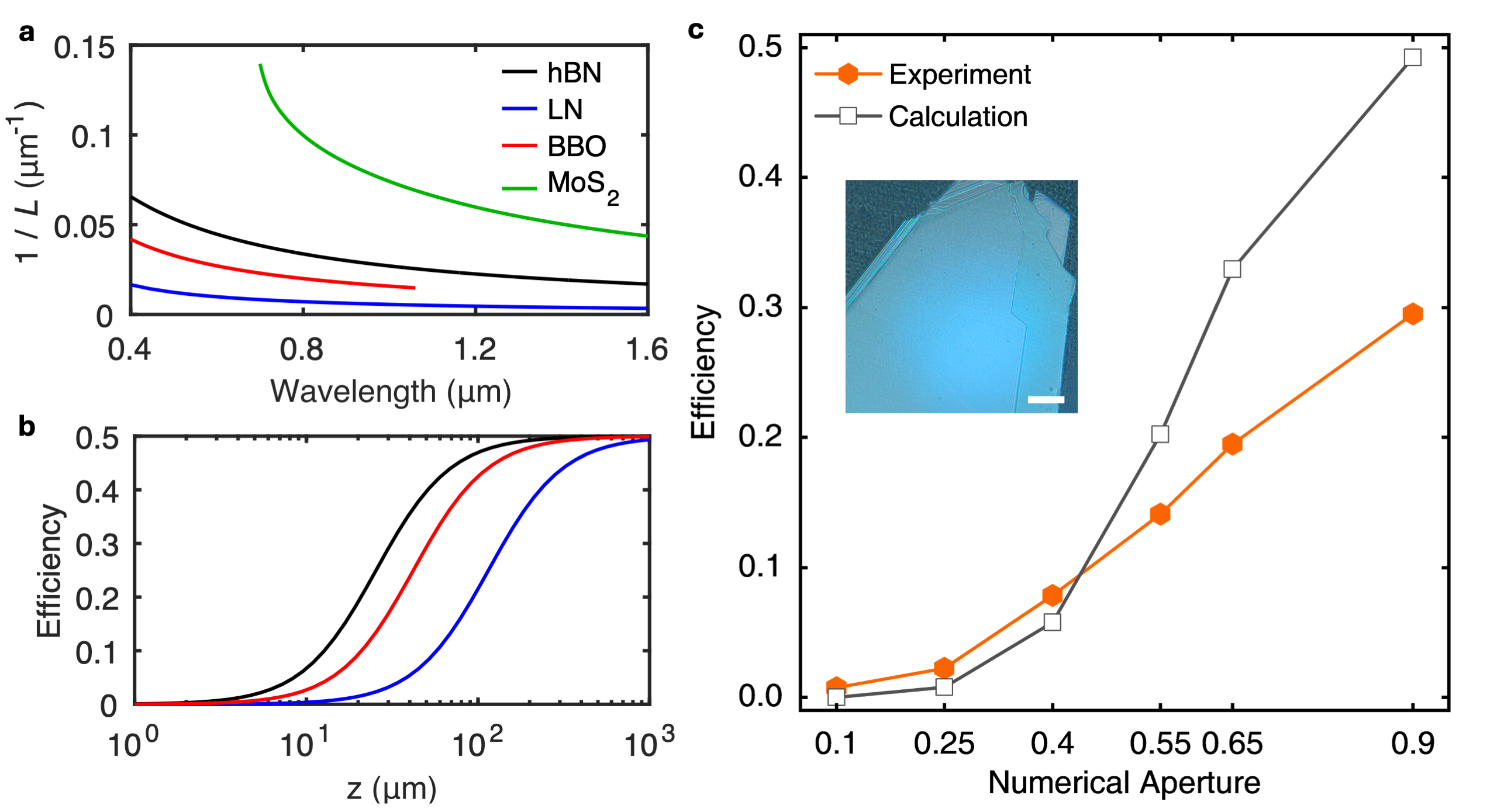}
    \caption{Spin-orbit conversion efficiency of anisotropic materials. (a) Inverses of anisotropic diffraction lengths ($1/L$) of anisotropic materials (hBN, BBO, LN, and MoS$_2$) by wavelength. (b) Analytically calculated spin-orbit conversion efficiencies of the anisotropic materials by the propagation length at $\lambda$=594 nm. (c) Analytically calculated conversion efficiency (black line) and experimental data (orange line) of an 8-\unit{\micro\metre}-thick hBN vortex generator at $\lambda$=594 nm. The inset shows an optical microscope (OM) image of the 8-\unit{\micro\metre}-thick hBN crystal used for the vortex generation. The scale bar is 50 \unit{\micro\metre}.}
    \label{fig:fig4}
\end{figure}

Figure~\ref{fig:fig4}a shows $1/L$ values of the various anisotropic materials (hBN, LN, BBO, and MoS$_2$) depending on the wavelength, calculated by using the dispersion data from the literature~\cite{rah2019optical, zelmon1997infrared, eimerl1987optical, ermolaev2021giant}. MoS$_2$ shows the largest $1/L$ value in the infrared and visible ranges, followed by hBN. This is due to their large anisotropy originating from the vdW structure. Figure~\ref{fig:fig4}b shows the conversion efficiencies achieved by Eq.~\ref{eqn:eq5} with respect to the propagation length when the CPL beam with $\lambda=594$ nm is focused by the 0.4 NA objective lens and propagates through the extraordinary axis of the anisotropic materials.  The half size of the beam waist, $w_0$, is approximated to be 0.62 \unit{\micro\metre}, using the equation 0.42$\lambda$/NA, an equation adopted for the Gaussian beam width estimation~\cite{zhang2007gaussian}. Note that the conversion efficiency of hBN is the highest among three anisotropic materials when they are under the same propagation length at $\lambda=594$ nm.

To compare the analytically predicted values with the experimental results, we conduct experiments to measure conversion efficiencies, defined as follows:
\begin{equation}
    \text{Conversion efficiency} = \frac{\text{Spin-converted output power}}{\text{Total output power}} = \frac{P_\text{PS2} / T_\text{LP}}{P_\text{PS1}}.
    \label{eqn:eq6}
\end{equation} 
Here, $P_\text{PS1}$ and $P_\text{PS2}$ indicate the power measured by photodiode power sensors at the position PS1 and PS2 in the setup schematic shown in Fig.~\ref{fig:fig3}a, respectively. $T_\text{LP}$ is the transmission efficiency of the linear polarizer introduced to compensate the reduced intensity due to reflection or absorption. $P_\text{PS1}$ represents the total output power from the vortex generator, and $P_\text{PS2}/T_\text{LP}$ refers to the output power of only the spin-converted beam. Figure~\ref{fig:fig4}c displays the experimental (orange line with hexagonal data points) and calculated (black line with square data points) conversion efficiencies of the 8-\unit{\micro\metre}-thick-hBN vortex generator with respect to the NA of OBL1. We set the NA of OBL2 to match that of OBL1, as outlined in Tab.~\ref{tab:tabS1} in the appendix, to minimize the loss of transmission power after the propagation through the sample. As theoretically predicted, the conversion efficiency increases with a higher NA because lenses with a higher NA provide a smaller beam waist. The highest conversion efficiency measured from hBN flakes is 0.30 at NA$=$0.9. However, this value is significantly lower than the analytic prediction, which we attribute to angle-dependent reflectance at the interfaces of the hBN crystal. As shown in Eq.~\ref{eqn:eq4}, the spin-orbit conversion rate by the propagation length increases with the amount of the transverse wavevector, which is proportional to the incidence angle. 
\begin{figure}
    \centering
    \includegraphics[width=1.0\linewidth]{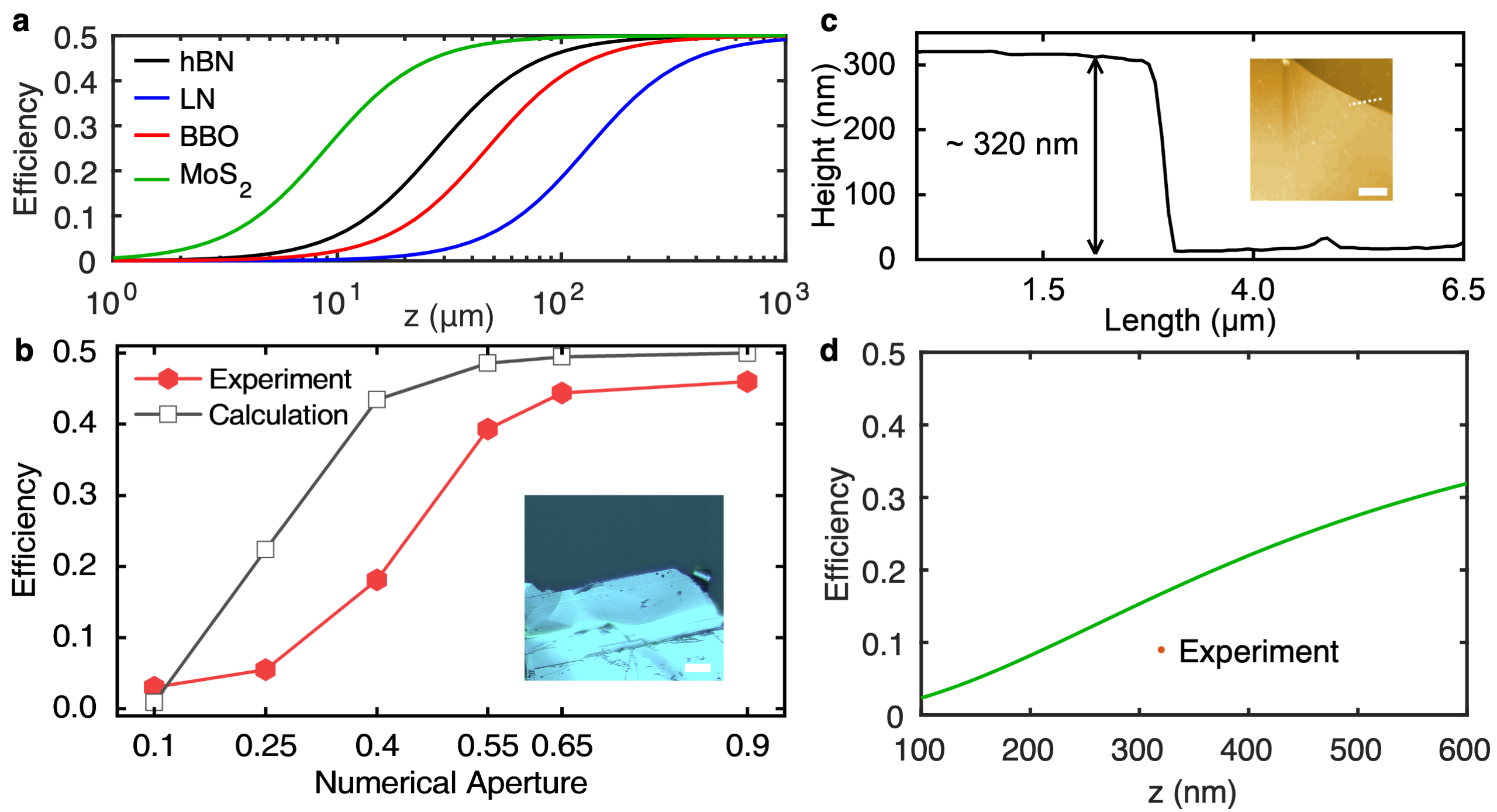}
    \caption{Spin-orbit conversion efficiency of MoS$_2$. (a) Calculated conversion efficiency of anisotropic materials at $\lambda$ = 750 nm when the NA is 0.4. (b) Conversion efficiency of bulk MoS$_2$ with respect to the NA when the thickness of MoS$_2$ is 26 \unit{\micro\metre}. The inset shows the OM image of a 26-\unit{\micro\metre}-thick MoS$_2$ crystal. The scale bar is 50 \unit{\micro\metre}. (c) The atomic force microscopy (AFM) scan shows cross-sections of an exfoliated MoS$_2$ flake along the white dashed line in the inset image. The inset image displays the AFM profile of the flake on the glass substrate. The scale bar is 4 \unit{\micro\metre}. (d) Calculated (green solid line) and experimental (red dot) efficiency of $\sim$ 320 nm MoS$_2$ crystal when the NA is 0.9 at $\lambda$ = 750 nm.}
    \label{fig:fig5}
\end{figure}

The conversion efficiency can be further enhanced by utilizing vdW materials with greater birefringence. MoS$_2$, one of the transition metal dichalcogenides, has giant birefringence and is transparent in the infrared region. Thus, it has excellent advantages in achieving high conversion efficiency in a compact size. Figure~\ref{fig:fig5}a shows the calculated conversion efficiencies with respect to the propagation length when MoS$_2$ and other anisotropic materials are applied as a vortex generator. The wavelength is 750 nm, and the minimum beam waist is 0.62 \unit{\micro\metre}, which corresponds to the beam focused by an objective lens of 0.4 NA. Owing to its large birefringence, the MoS$_2$-based vortex generator displays exceedingly higher conversion efficiency than the other anisotropic materials under the same propagation length.
We measured the conversion efficiency of the MoS$_2$-based vortex generator using the same setup of the hBN vortex generator in Fig.~\ref{fig:fig3}a except for the laser, which is replaced by one with a wavelength of 750 nm. Figure~\ref{fig:fig5}b shows the conversion efficiency of the vortex generator using the 26-\unit{\micro\metre}-thick MoS$_2$ crystal depending on the NA of OBL 1, compared with the calculation. The measured conversion efficiency increases with the NA and approaches 0.46 at NA = 0.9 but is lower than the calculation at all NA values.

We achieved vortex generation on a sub-wavelength scale using a MoS$_2$ flake. First, we exfoliated the MoS$_2$ flake from a bulk MoS$_2$ crystal and positioned it on a cover glass as shown in the optical microscope image in Fig.~\ref{fig:fig5}c. We subsequently measured the thickness of the flake with an atomic-force-microscope (AFM) and verified that it was 320 nm. Furthermore, We measured the conversion efficiency of the MoS$_2$ vortex generator and plotted it on Fig.~\ref{fig:fig5}d.  The vortex generator demonstrated a conversion efficiency of 0.09 at NA = 0.9, which is lower than the calculated value.

\begin{figure}
    \centering
    \includegraphics[width=1.0\linewidth]{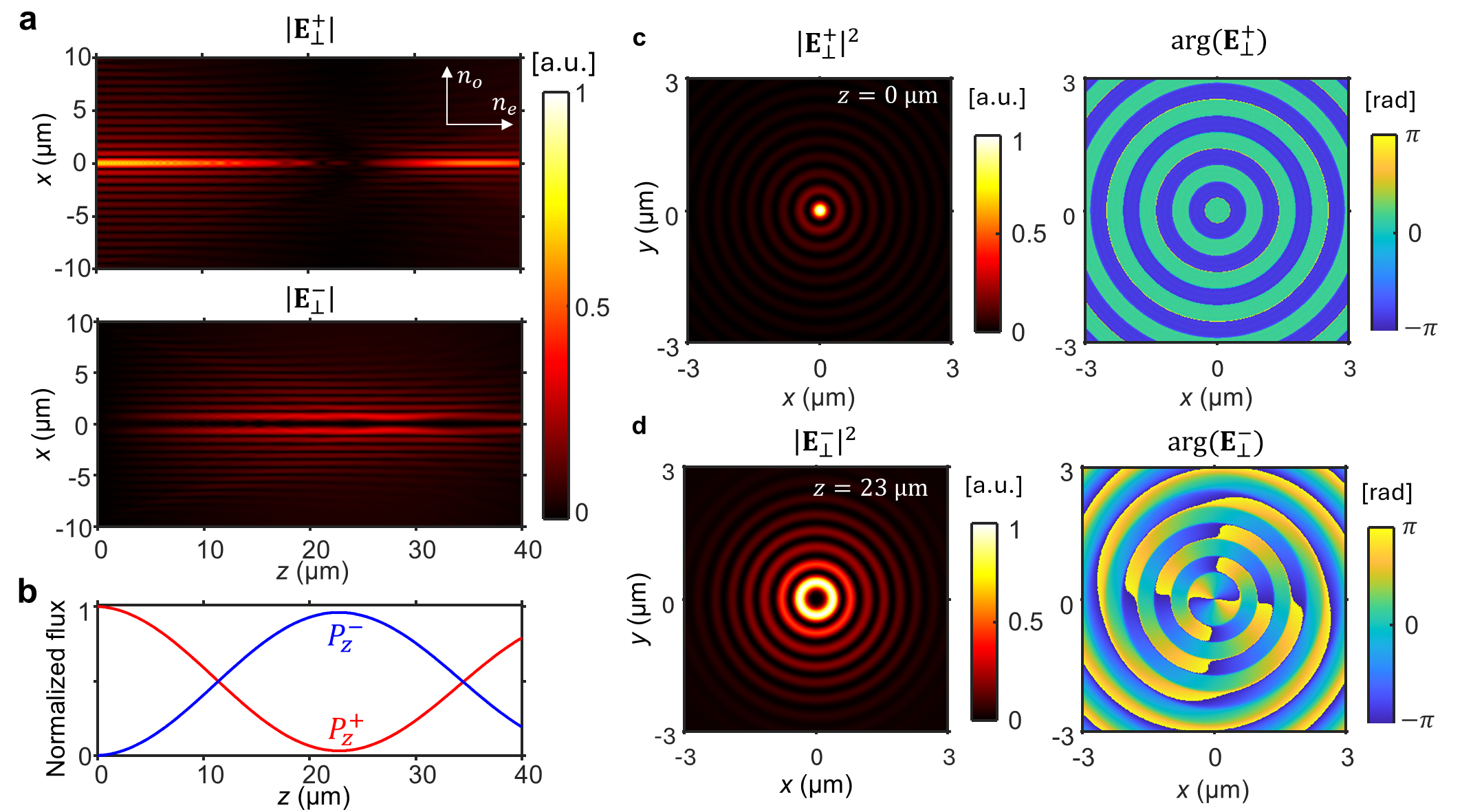}
    \caption{Spin-orbit coupling of a Bessel beam. (a) Amplitude distribution of a circularly polarized Bessel beam propagating in the hBN crystal along the extraordinary axis when $\lambda$ = 594 nm. The $|E^+|$ and $|E^-|$ indicate the absolute amplitudes of the LCP and RCP electric fields, respectively, which are normalized by the maximum amplitudes of the LCP wave. (b) $z$-directional powers of the LCP ($P_z^+$) and RCP ($P_z^-$) waves. (c) Intensity and phase profiles of the LCP wave at $z=0$ \unit{\micro\metre}. (d) and of the RCP wave at $z=23$ \unit{\micro\metre}.}
    \label{fig:fig6}
\end{figure}

The use of the Bessel beam can overcome the Gaussian beam's maximum conversion efficiency limit of 0.5, enabling the near-unity conversion. Figure~\ref{fig:fig6}a shows the amplitude profiles of the LCP and RCP fields of the Bessel beam propagating along the extraordinary axis of the hBN crystal at $\lambda$=594 nm. The LCP Bessel beam is generated at $z=0$ \unit{\micro\metre} with a diameter of 10 \unit{\micro\metre} and a transverse wavevector $k_t$ of 0.4 $k_0$. As $z$ increases from 0 to 23 \unit{\micro\metre}, the LCP field almost disappears, and the RCP field intensity rises. Figure~\ref{fig:fig6}b shows the normalized $z$-directional powers of the LCP and RCP waves. Those powers oscillate sinusoidally, and the RCP power gets the highest value of 0.96 at $z=23$ \unit{\micro\metre}. Figures~\ref{fig:fig6}c and d show the intensity and phase profiles of the LCP and RCP components, respectively, revealing an RCP Bessel beam with the OAM mode of $+2$ at $z=23$ \unit{\micro\metre}.

The spin-orbit conversion of the Bessel beam can be analytically derived by Eq.~\ref{eqn:eq4}. The ideal LCP amplitudes of Bessel beam in the Fourier space are expressed by a delta function; $\tilde{U}_+(\textbf{k}_\perp,0)$ = $\tilde{U}_0\delta(k_\perp-k_t)$ and $\tilde{U}_-(\textbf{k}_\perp,0) = 0$, where $\tilde{U}_0$ is an arbitrary amplitude. Thus, the conversion efficiency of the Bessel beam is given by:
\begin{equation}
    \eta_B = \left| t_{-+}(k_t,z) \right|^2  = \frac{1}{4}\left|\exp(ik_{ez}z)-\exp(ik_{oz}z)\right|^2 = \sin^2\left(\frac{k_{oz}-k_{ez}}{2}z\right).
\end{equation}
In the case of the hBN crystal, $\eta_B$ approaches 1 at $z=23$ \unit{\micro\metre} as shown in the simulation. We can further reduce the propagation length for the near-unity spin-orbit conversion by increasing the transverse wavevector $k_t$ or by using a medium with a larger birefringence.

\section{Conclusion}
In conclusion, we present the generation of the optical vortex beam by leveraging the spin-orbit coupling in vdW materials, eliminating the need for fabrication processes. By utilizing a hBN crystal, we produced a doughnut-shaped beam profile with the topological charge of $\pm$2 through the conversion of circular polarization. The conversion efficiencies of the vortex generators were measured, revealing that the 8-\unit{\micro\metre}-thick hBN crystal and the 23-\unit{\micro\metre}-thick MoS$_2$ crystal achieved maximum conversion efficiencies of 0.30 and 0.46, respectively.  Furthermore, We demonstrated spin-orbit coupling on the sub-wavelength scale, achieving the conversion efficiency of 0.09 using the 320-nm-thick MoS$_2$ crystal flake. In addition, our numerical simulations further suggest that near-unity conversion efficiency can be realized in vdW vortex generators by employing Bessel beams. These ultra-compact, fabrication-free vortex beam generators represent a potential in nanophotonics, particularly in applications involving orbital angular momentum.

\setcounter{figure}{0}
\renewcommand{\thefigure}{S\arabic{figure}}
\renewcommand{\thetable}{S\arabic{table}}

\section{Appendix}
\noindent\textbf{Simulation method}

For the simulations involving Gaussian beam incidence, we utilize a cylindrical symmetry with a simulation domain of 22.1 \unit{\micro\metre} in length and 11 \unit{\micro\metre} in radius. The grid spacing is set to 10 nm along both the $z$-axis and the radial coordinate $r_\perp$ of the cylindrical coordinate system. The boundaries of the simulation space are surrounded by 1-\unit{\micro\metre}-thick perfect matching layers (PMLs) to absorb all outgoing waves. A source plane is positioned 20 nm to the right from the left boundary ($-z$). The entire simulation space is filled with hBN, characterized by refractive indices of $n_o=2.15$ and $n_e=1.86$. The simulation requires approximately 5 minutes on a workstation equipped with a 64-core CPU (AMD Ryzen Threadripper 5995WX). For the Bessel beam simulations, the same setup is used, except the radius is increased to 21 \unit{\micro\metre}, and the radius of the source plane is set to 10 \unit{\micro\metre}. This simulation takes approximately 10 minutes on the same workstation.

\begin{figure}
    \centering
    \includegraphics[width=1.0\linewidth]{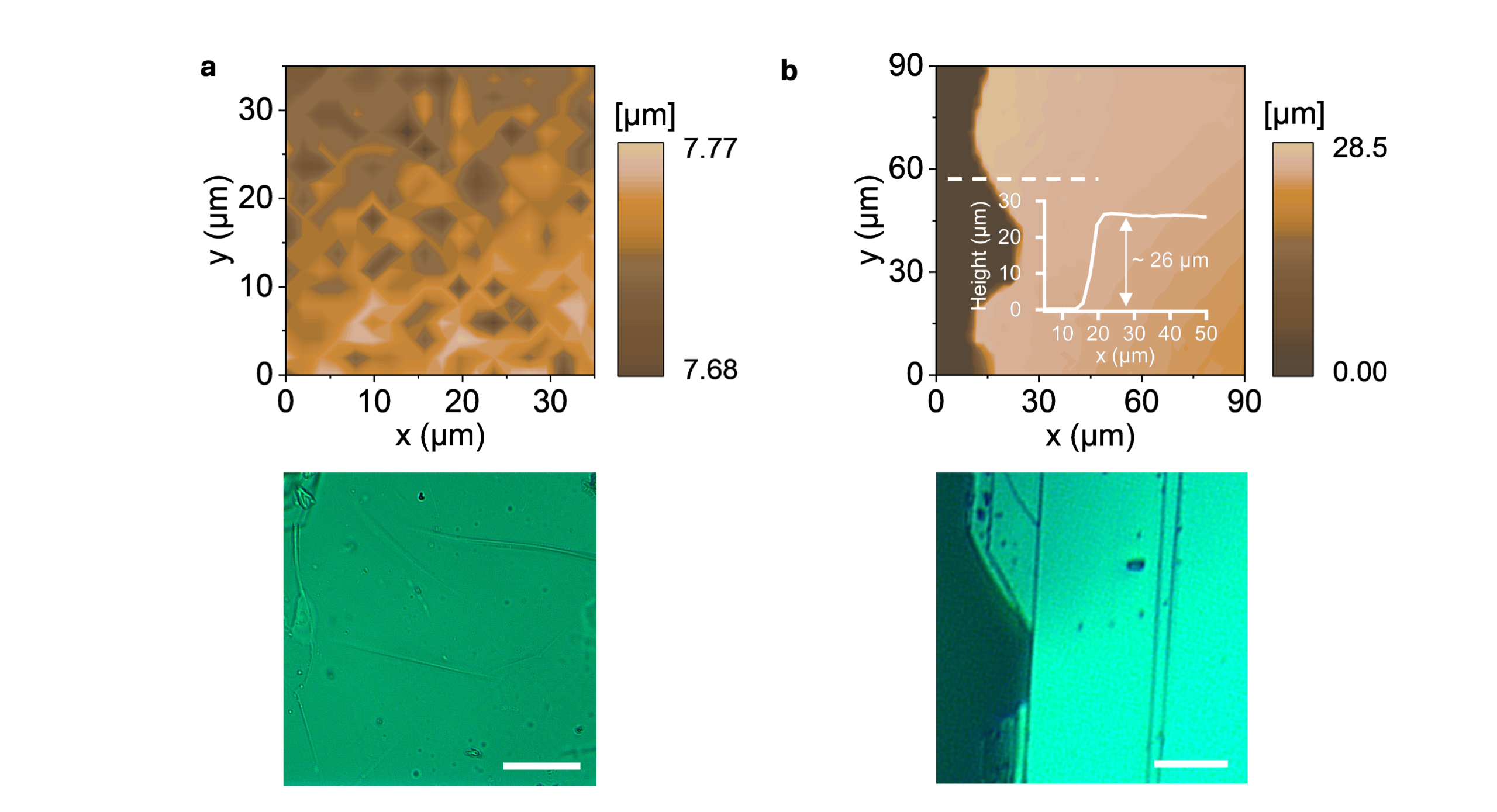}
    \caption{Thickness profiles and OM images of the hBN and MoS$_2$ crystals. (a) (top panel) A thickness profile of the hBN crystal measured by a profilometer. (bottom panel) An OM image of the area where the thickness profile is measured. (b) (top panel) A thickness profile at the edge of the MoS$_2$ crystal. The inset graph shows a one-dimensional cross-section profile of the white dashed line marked in the thickness profile. (bottom panel) An OM image of the area where the profile is measured. The scale bars of both OM images are 20 \unit{\micro\metre}. Note that the profilometer can obtain accurate thickness even in areas without a crystal edge because it uses optical interference for measurements.}
    \label{fig:figS1}
\end{figure}

\noindent\textbf{Experiments}

The hBN and MoS$_2$ crystals are purchased from 2D Semiconductors, and Bruker Contour GT-I profilometer using a green light source is applied to measure the thicknesses of the crystals. In addition, the MoS$_2$ flake is mechanically exfoliated from the MoS$_2$ crystal by using the Kapton tape and transferred to a 0.1-mm-thick cover glass. Moreover, an AFM equipment of Asylum Research Cypher is utilized to measure the thickness of the MoS$_2$ flake. Coherent OBIS LS 594 nm 60 mW laser is applied for the vortex generation in the hBN crystal.  Furthermore, 750 nm laser, obtained by filtering the 10-nm-wide band from the SuperK Fianium supercontinuum laser source, is used for the MoS$_2$ crystal and flake. The beam intensity profile is captured by YW500 camera from ShenZhen YangWang Technology Co.

\begin{table}
    \centering
    \begin{tabular}{|c|c|}
    \hline
    OBL 1 & OBL 2 \\
    \hline
    0.1 & 0.25 \\
    \hline
    0.25 & 0.28 \\
    \hline
    0.4 & 0.42 \\
    \hline
    0.55 & 0.42 \\
    \hline
    0.65 & 0.55 \\
    \hline
    0.9 & 0.9 \\
    \hline
    \end{tabular}
    \caption{NA of OBL 1 and following OBL 2 implemented in transmission setup in efficiency measurement experiment}
    \label{tab:tabS1}
\end{table}

\printbibliography
\end{document}